\documentclass[superscriptaddress,twocolumn,english,floatfix,aps,pra,amsmath,amssymb,longbibliography]{revtex4-2}
\usepackage[T1]{fontenc}
\usepackage[utf8]{inputenc}
\usepackage{babel}
\usepackage{comment}
\usepackage{color}
\usepackage{times}
\usepackage{epsfig}
\usepackage{subfigure}
\usepackage{amsmath}
\usepackage{amssymb}
\usepackage{graphicx}
\usepackage[colorlinks=true]{hyperref}
\hypersetup{citecolor=blue,linkcolor=blue,urlcolor=blue}

\begin{document}

\title{Exact Solution of a Time-Dependent Quantum Harmonic Oscillator with Two Frequency Jumps via the Lewis-Riesenfeld Dynamical Invariant Method}

\author{Stanley S. Coelho}
\email{stanley.coelho@icen.ufpa.br}
\affiliation{Faculdade de F\'{i}sica, Universidade Federal do Par\'{a}, 66075-110, Bel\'{e}m, Par\'{a}, Brazil}
	
\author{Lucas Queiroz}
\email{lucas.silva@icen.ufpa.br}
\affiliation{Faculdade de F\'{i}sica, Universidade Federal do Par\'{a}, 66075-110, Bel\'{e}m, Par\'{a}, Brazil}
	
\author{Danilo T. Alves}
\email{danilo@ufpa.br}
\affiliation{Faculdade de F\'{i}sica, Universidade Federal do Par\'{a}, 66075-110, Bel\'{e}m, Par\'{a}, Brazil}
\affiliation{Centro de F\'{i}sica, Universidade do Minho, P-4710-057, Braga, Portugal}
	
\date{\today}
	
\begin{abstract}
Harmonic oscillators with multiple abrupt jumps in their frequencies have been investigated by several authors during the last decades.
We investigate the dynamics of a quantum harmonic oscillator with initial frequency $\omega_0$, that undergoes a sudden jump to a frequency $\omega_1$ and, after a certain time interval, suddenly returns to its initial frequency.
Using the Lewis-Riesenfeld method of dynamical invariants, 
we present expressions for the mean energy value, the mean number of excitations, and the transition probabilities, considering the initial state different from the fundamental.
We show that the mean energy of the oscillator, after the jumps, is equal or greater than the one before the jumps, even when $\omega_1<\omega_0$.
We also show that, for particular values of the time interval between the jumps, the oscillator returns to the same initial state. 
\end{abstract}

\maketitle

%
\section{Introduction}
\label{sec:introduction}
The quantum harmonic oscillator potential with time-dependent parameters is relevant in modeling several problems in physics and has been investigated \cite{Husimi-PTP-1953-II,Lewis-PRL-1967,Lewis-JMP-1968,Lewis-JMP-1969,Pedrosa-PRA-1997,Ciftja-JPA-1999,Guasti-JPA-2003,Pedrosa-PRL-2009,Pedrosa-PRA-2011}.
For example, the interaction between a spinless charged quantum particle and a time-dependent external classical electromagnetic field can be studied through a harmonic potential whose frequency depends explicitly on time \cite{Lewis-JMP-1969, Dodonov-PLA-1994, Xu-CP-1999, Aguiar-JMP-2016, Dodonov-JRLR-2018}, 
and this is used to model the quantum motion of this particle in a trap \cite{Brown-PRL-1991, Agarwal-PRL-1991, Mihalcea-PS-2009, Aguiar-IJMS-2016, Menicucci-PRA-2007, Pedrosa-BJP-2021}.
In the context of quantum electrodynamics, this potential is useful, for instance, to describe the free electromagnetic field in nonstationary media \cite{Pedrosa-PRL-2009,Pedrosa-PRA-2011,Choi-PRA-2010}. 
In the context of shortcuts to adiabaticity, time-dependent quantum oscillators have also been considered \cite{Salamon-PCCP-2009, Schaff-PRA-2010, Chen-PRL-2010, Stefanatos-PRA-2010, Dupays-PRR-2021,Tibaduiza-JPB-2021}.
Other applications are found in relativistic quantum mechanics, quantum field theory, dynamical Casimir effect and gravitation \cite{Landim-PRA-2000,Gao-PRA-1998,Dodonov-PRA-1996,Dodonov-PLA-1990,Pedrosa-IJMP-2004,Pedrosa-PRD-2004,Greenwood-IJMP-2015}. 

A particular case of a quantum harmonic oscillator with time-dependent parameters, that shows sudden frequency jumps, is investigated, for instance, in Refs. \cite{Janszky-OC-1986,Janszky-PRA-1992,Kiss-PRA-1994,Cessa-PLA-2003,Salamon-PCCP-2009,Stefanatos-PRA-2010,Chen-PRL-2010,Stefanatos-SJCO-2017,Stefanatos-TAC-2017,Tibaduiza-BJP-2020,Tibaduiza-PS-2020}.
Under such jumps (or any time dependence in the parameters), 
a classical oscillator in its ground state remains in the same state, whereas
a quantum oscillator can become excited \cite{Janszky-PRA-1992}.
Moreover, the wave functions of quantum harmonic oscillators with time-dependent parameters describe squeezed states \cite{Pedrosa-PRA-1997,Pedrosa-PRA-1997-singular-perturbation,Pedrosa-PRA-2011}.
For example, a sudden change in the oscillation frequency of $^{\text{85}}\text{Rb}$ atoms in the vibrational fundamental state of a one-dimensional optical lattice generates squeezed states \cite{Xin-PRL-2021}.
The description of squeezed states is relevant, for instance, in the implementation of schemes for noise minimization in quantum sensors, which increases their sensitivity (see, for instance, Ref. \cite{Wolf-Nature-2019} and references therein).
Subtle points involving the squeezed states for the model of two frequency jumps were investigated, for instance, by Tibaduiza {\textit{et al.}} in Ref. \cite{Tibaduiza-BJP-2020}, where the solution for this case was obtained via algebraic method.

In the present paper, we investigate the dynamics of a quantum harmonic oscillator with initial frequency $\omega_0$, that undergoes a sudden jump to a frequency $\omega_1$ and, after a certain time interval, suddenly returns to its initial frequency.
Instead of using the algebraic method used in Ref. \cite{Tibaduiza-BJP-2020}, here we use the Lewis-Riesenfeld (LR) method of dynamical invariants.
The LR method \cite{Lewis-PRL-1967,Lewis-JMP-1968,Lewis-JMP-1969} enables the calculation of the exact wave function of a system subjected, for instance, to a harmonic oscillator potential with time-dependent parameters, such as mass and frequency
\cite{Pedrosa-PRA-1997,Pedrosa-PRA-1997-singular-perturbation,Choi-IJMPB-2004}. 
Using this method, we show that the results for the squeeze parameters, the quantum fluctuations of the position and momentum operators, and the probability amplitude of a transition from the fundamental state to an arbitrary energy eigenstate coincide with those found in Ref. \cite{Tibaduiza-BJP-2020}.
In addition, using the same LR method, we also obtain expressions for the mean energy value and for the mean number of excitations (which were not calculated in Ref. \cite{Tibaduiza-BJP-2020}), and for the transition probabilities considering the initial state different from the fundamental (which generalizes the formula found in Ref. \cite{Tibaduiza-BJP-2020}).

The paper is organized as follows. 
In Sec. \ref{sec:aplicação do método LR ao oscilador}, we review some results on the application of the LR method to the quantum harmonic oscillator with time-dependent frequency. 
In Sec. \ref{sec:fluctuations, squeezed states and excitations}, we define the squeezing parameters and, from these and the oscillator wave function obtained via the LR method, we determine the quantum fluctuations of the position, momentum and Hamiltonian operators, the mean number of excitations, and the transition probabilities between different states.  
In Sec. \ref{sec:model}, we apply the results of previous sections to the model of Ref. \cite{Tibaduiza-BJP-2020}, and analyze their physical implications.
In Sec. \ref{sec:final}, we present our final remarks.

\section{Analytical Method}
\label{sec:review}

\subsection{The Wave Function of the Harmonic Oscillator via Lewis-Riesenfeld Method}
\label{sec:aplicação do método LR ao oscilador}

Let us consider the one-dimensional Schrödinger equation for a system whose Hamiltonian $ \hat{H}(t) $ explicitly depends on time \cite{Sakurai-Quantum-Mechanics-2021, Griffiths-Quantum-Mechanics-2005, Tannoudji-Quantum-Mechanics-2019},
\begin{eqnarray}\label{eq:equação de Schrodinger}
	i\hbar\frac{\partial\Psi(x,t)}{\partial t}=\hat{H}(t)\Psi(x,t).
\end{eqnarray}
According to the LR method \cite{Lewis-PRL-1967,Lewis-JMP-1968,Lewis-JMP-1969,Pedrosa-PRA-1997,Pedrosa-PRA-1997-singular-perturbation,Pedrosa-PRA-2011}, given an invariant Hermitian operator $ \hat{I}(t)$, which satisfies 
\begin{eqnarray}\label{eq:equação de Heisenberg}
	\frac{\partial\hat{I}(t)}{\partial t}+\frac{1}{i\hbar}\bigl[\hat{I}(t),\hat{H}(t)\bigr]=0,
\end{eqnarray}
a particular solution $\Psi_n(x,t)$ of Eq. \eqref{eq:equação de Schrodinger} is 
\begin{eqnarray}
	\label{eq:Psi-n-arb}
	\Psi_{n}(x,t)=\exp\left[i\alpha_{n}(t)\right]\Phi_{n}(x,t),
\end{eqnarray}
in which $ \Phi_{n}(x,t) $ are the eigenfunctions of $ \hat{I}(t) $, found from
\begin{eqnarray}
	\hat{I}(t)\Phi_{n}(x,t)=\lambda_{n}\Phi_{n}(x,t),
\end{eqnarray}
with $ \lambda_n $ being time independent eigenvalues of $ \hat{I}(t)$, and $\alpha_{n}(t)$ phase functions, obtained from the equation 
\begin{eqnarray}
	\label{eq:alpha-n}	
	\frac{d\alpha_{n}(t)}{dt}=\int_{-\infty}^{+\infty}dx\,\Phi_{n}^{*}(x,t)\biggl[i\frac{\partial}{\partial t}-\frac{1}{\hbar}\hat{H}(t)\biggr]\Phi_{n}(x,t).
\end{eqnarray}
The general solution $ \Psi(x,t) $ of Eq. \eqref{eq:equação de Schrodinger} is 
\begin{equation}
	\Psi(x,t)=\sum_{n=0}^{\infty}C_{n}\Psi_{n}(x,t),
\end{equation}
where the time independent coefficients $ C_n $ depend only on the initial conditions.

Specifically, for a time-dependent one-dimensional harmonic oscillator with mass $ m_0 $, whose time-dependence is contained purely in its oscillation frequency $ \omega(t) $, the Hamiltonian is given by 
\begin{eqnarray}\label{eq:hamiltoniano do oscilador}
	\hat{H}(t)=\frac{\hat{p}^{2}}{2m_{0}}+\frac{1}{2}m_{0}\,\omega(t)^{2}\hat{x}^{2},
\end{eqnarray}
where $ \hat{x} $ and $ \hat{p} $ are position and momentum operators, respectively, with $ \left[\hat{x},\hat{p}\right]=i\hbar $.
An operator $ \hat{I}(t) $ associated with Eq. \eqref{eq:hamiltoniano do oscilador} is \cite{Lewis-JMP-1969,Pedrosa-PRA-1997,Pedrosa-PRA-1997-singular-perturbation,Pedrosa-PRA-2011}
\begin{eqnarray}
	\label{eq:I-oscilador}
	\hat{I}(t)=\frac{1}{2}\left\{ \left[\frac{\hat{x}}{\rho(t)}\right]^{2}+\left[\rho(t)\hat{p}-m_{0}\dot{\rho}(t)\hat{x}\right]^{2}\right\},
\end{eqnarray}
wherein $ \rho(t) $ is a real parameter which is solution of the Ermakov-Pinney equation \cite{Prykarpatskyy-JMS-2018, Pinney-PAMS-1950,Lima-JMO-2009,Carinena-IJGMMP-2009} 
\begin{eqnarray}
	\label{eq:equação de Ermakov-Pinney}
	\ddot{\rho}(t)+\omega(t)^{2}\rho(t)=\frac{1}{m_{0}^{2}\rho(t)^{3}}.
\end{eqnarray}
The eigenfunctions of $ \hat{I}(t) $, given by Eq. \eqref{eq:I-oscilador}, are
\begin{eqnarray}
	\label{eq:phi_n}
	\Phi_{n}(x,t)=\frac{1}{\sqrt{2^{n}n!}}\Phi_{0}(x,t)H_{n}\left[\frac{x}{\hbar^{\frac{1}{2}}\rho(t)}\right],
\end{eqnarray}
where $ H_{n}$ are the Hermite polynomials of order $ n $ \cite{Arfken-Mathematical-Physics-2003}
, $ \lambda_{n}=\left(n+1/2\right)\hbar $, and
\begin{eqnarray}
	\nonumber
	\Phi_{0}(x,t)=\biggl[\frac{1}{\pi\hbar\rho(t)^{2}}\biggr]^{\frac{1}{4}}\exp\biggl\{\frac{im_{0}}{2\hbar}\biggl[\frac{\dot{\rho}(t)}{\rho(t)}+\frac{i}{m_{0}\rho(t)^{2}}\biggr]x^{2}\biggr\}.\\
	\label{eq:phi0}
\end{eqnarray}
From Eq. \eqref{eq:alpha-n}, the functions $ \alpha_{n}(t) $ are given by
\begin{eqnarray}
	\label{eq:alpha_n}
	\alpha_{n}(t)=-\frac{1}{m_{0}}\left(n+\frac{1}{2}\right)\int_{0}^{t}\frac{dt^{\prime}}{\rho\left(t^{\prime}\right)^{2}}.
\end{eqnarray}
Thus, from Eqs. \eqref{eq:Psi-n-arb}, \eqref{eq:phi_n} and \eqref{eq:alpha_n}, the wave function $ \Psi_{n}(x,t) $ associated with the Hamiltonian \eqref{eq:hamiltoniano do oscilador} is 
\begin{eqnarray}
	\nonumber
	\label{eq:função de onda do oscilador}
	\Psi_{n}(x,t)=\frac{1}{\sqrt{2^{n}n!}}\exp\left[-\frac{i}{m_{0}}\left(n+\frac{1}{2}\right)\int_{0}^{t}\frac{dt^{\prime}}{\rho(t^{\prime})^{2}}\right]\\
	\times\Phi_{0}(x,t)H_{n}\left[\frac{x}{\hbar^{\frac{1}{2}}\rho(t)}\right].
\end{eqnarray}
For the case in which the frequency is always constant [$\omega(t)=\omega_{0}$], 
the solution of Eq. \eqref{eq:equação de Ermakov-Pinney}
is $\rho(t)=\rho_0$, where \cite{Lewis-JMP-1969,Pedrosa-PRA-1997-singular-perturbation,Ciftja-JPA-1999}
\begin{eqnarray}
	\label{eq:rho0}
	\rho_{0}=\frac{1}{\sqrt{m_{0}\omega_{0}}}.
\end{eqnarray}
Therefore, Eq. \eqref{eq:função de onda do oscilador} falls back to the wave function of a harmonic oscillator with time independent mass and frequency, 
$\Psi_{n}^{(0)}(x,t)$, given by \cite{Sakurai-Quantum-Mechanics-2021,Griffiths-Quantum-Mechanics-2005,Tannoudji-Quantum-Mechanics-2019}
\begin{eqnarray}
	\nonumber
	\Psi_{n}^{(0)}(x,t)=\frac{1}{\sqrt{2^{n}n!}}\Bigl(\frac{m_{0}\omega_{0}}{\pi\hbar}\Bigr)^{\frac{1}{4}}\exp\biggl[-i\left(n+\frac{1}{2}\right)\omega_{0}t\biggr]\\
	\times\exp\biggl(-\frac{m_{0}\omega_{0}x^{2}}{2\hbar}\biggr)H_{n}\biggl[\left(\frac{m_{0}\omega_{0}}{\hbar}\right)^{\frac{1}{2}}x\biggr].
	\label{eq:Psi-0}
\end{eqnarray}
%

\subsection{Squeeze Parameters, Quantum Fluctuations, Mean Number of Excitations, and Transition Probability}
\label{sec:fluctuations, squeezed states and excitations}

As discussed in Refs. \cite{Pedrosa-PRD-1987,Pedrosa-PRA-1997,Pedrosa-PRA-1997-singular-perturbation,Pedrosa-PRA-2011}, the quantum states of the time-dependent oscillator, characterized by the wave function $ \Psi_{n}(x,t) $ [Eq. \eqref{eq:função de onda do oscilador}], are squeezed.
Thus, we can define the squeeze parameter $r(t)$ and the squeeze phase $\phi(t)$, which
specify the squeezed state, in terms of the parameter $\rho(t)$ \cite{Daneshmand-CTP-2017}:
\begin{eqnarray}
	\nonumber
	r(t)=\cosh^{-1}\Bigl\{\left(4m_{0}\omega_{0}\right)^{-\frac{1}{2}}\bigl[m_{0}^{2}\dot{\rho}(t)^{2}+\rho(t)^{-2}\\
	\label{eq:r}
	+2m_{0}\omega_{0}+m_{0}^{2}\omega_{0}^{2}\rho(t)^{2}\bigr]^{\frac{1}{2}}\Bigr\},\\
    \label{eq:phi}
	\phi(t)=\cos^{-1}\left\{ \frac{1+m_{0}\omega_{0}\rho(t)^{2}-2\cosh^{2}[r(t)]}{2\sinh[r(t)]\cosh[r(t)]}\right\},
\end{eqnarray}
with $r(t)\geq 0 $ and $ 0\leq\phi(t)\leq 2\pi$.
From Eq. \eqref{eq:função de onda do oscilador}, one can also obtain the expected value of a given observable $ \hat{O}(t) $ in the state $ \Psi_{n}(x,t) $, as
\begin{eqnarray}\label{eq:<O>}
	\langle\hat{O}(t)\rangle(n,t)=\int_{-\infty}^{+\infty}dx\,\Psi_{n}^{*}(x,t)\,\hat{O}(t)\,\Psi_{n}(x,t),
\end{eqnarray}
which, from Eqs. \eqref{eq:r} and \eqref{eq:phi}, can be written in terms of $r(t)$ and $\phi(t)$.
For the operators $ \hat{x} $ and $ \hat{p} $, one has \cite{Guerry-Quantum-Optics-2005}: 
\begin{eqnarray}
\langle\hat{x}\rangle(n,t)=\langle\hat{p}\rangle(n,t)=0,
\label{eq:<x>-<p>}
\end{eqnarray}
\begin{eqnarray}
\nonumber
\langle\hat{x}^{2}\rangle(n,t)=\left(n+\frac{1}{2}\right)\frac{\hbar}{m_{0}\omega_{0}}\bigl\{\cosh^{2}\left[r(t)\right]+\sinh^{2}\left[r(t)\right]\\
\nonumber
+\,2\sinh\left[r(t)\right]\cosh\left[r(t)\right]\cos\left[\phi(t)\right]\bigr\},\\
\label{eq:<x^2>}\\
\nonumber
\langle\hat{p}^{2}\rangle(n,t)=\left(n+\frac{1}{2}\right) m_{0}\omega_{0}\hbar\bigl\{\cosh^{2}\left[r(t)\right]+\sinh^{2}\left[r(t)\right]\\
\nonumber
-\,2\sinh\left[r(t)\right]\cosh\left[r(t)\right]\cos\left[\phi(t)\right]\bigr\},\\
\label{eq:<p^2>}
\end{eqnarray}
where it follows, from Eqs. \eqref{eq:hamiltoniano do oscilador}, \eqref{eq:<x^2>}, and \eqref{eq:<p^2>}, that
\begin{eqnarray}
		\label{eq:<H>}
	\langle\hat{H}(t)\rangle(n,t)=\frac{\langle\hat{p}^{2}\rangle(n,t)}{2m_{0}}+\frac{1}{2}m_{0}\,\omega(t)^{2}\langle\hat{x}^{2}\rangle(n,t).
\end{eqnarray}
From Eqs. \eqref{eq:<x>-<p>}-\eqref{eq:<p^2>}, one finds the variances of the operators $ \hat{x} $:
\begin{eqnarray}
\label{eq:Delta x^2}	
\langle[\Delta\hat{x}]^{2}\rangle(n,t)=\langle\hat{x}^{2}\rangle(n,t)-[\langle\hat{x}\rangle(n,t)]^{2},
\end{eqnarray}
and $ \hat{p} $:
\begin{eqnarray}
\label{eq:Delta p^2}
\langle[\Delta\hat{p}]^{2}\rangle(n,t)=\langle\hat{p}^{2}\rangle(n,t)-[\langle\hat{p}\rangle(n,t)]^{2},
\end{eqnarray}
which implies the uncertainty relationship
\begin{eqnarray}
\nonumber
\langle[\Delta\hat{x}]^{2}\rangle(n,t)\langle[\Delta\hat{p}]^{2}\rangle(n,t)\geq\biggl(n+\frac{1}{2}\biggr)^{2}\hbar^{2}\bigl\{\cosh^{4}[r(t)]\\
+\sinh^{4}[r(t)]-2\sinh^{2}[r(t)]\cosh^{2}[r(t)]\cos[2\phi(t)]\bigr\}.
\label{eq:incerteza}
\end{eqnarray}

Due to the time-dependence of the frequency, one can also determine the mean number of excitations $\langle\hat{N}\rangle(n,t)$ that a system, subjected to this potential, can undergo. 
This is given by \cite{Kim-PRA-1989,Marian-PRA-1992,Moeckel-AP-2009} 
\begin{eqnarray}
	\label{eq:<N>-2}
	\langle\hat{N}\rangle(n,t)=n+\left(2n+1\right)\sinh^{2}\left[r(t)\right].
\end{eqnarray}
For the fundamental state $ n=0 $, one finds $ 	\langle\hat{N}\rangle(0,t)=\sinh^{2}\left[r(t)\right] $, a result that agrees with Refs. \cite{Guerry-Quantum-Optics-2005,Greenwood-IJMP-2015} for vacuum squeezed states.
This means that, a system, even in the fundamental state, could be excited due to the temporal variations in its frequency.
The system subjected to the time-dependent harmonic potential can also make transitions between different states, since time-dependent potentials induce quantum systems to make transitions  \cite{Griffiths-Quantum-Mechanics-2005,Sakurai-Quantum-Mechanics-2021,Tannoudji-Quantum-Mechanics-2019}.
Let us consider that the system is initially at a stationary state $\Psi_{m}^{(0)}(x,t=0)$  with frequency $\omega_{0}$
and, due to a modification in its frequency from $\omega_{0}$ to $\omega(t)$, it evolves to a new state $ \Psi_{m}(x,t)$ [Eq. \eqref{eq:função de onda do oscilador}].
In this way, the probability to find the system in the state $\Psi_{n}^{(0)}(x,t)$ [Eq. \eqref{eq:Psi-0}], 
is given by \cite{Griffiths-Quantum-Mechanics-2005, Sakurai-Quantum-Mechanics-2021,Tannoudji-Quantum-Mechanics-2019}
\begin{eqnarray}
	{\cal P}(t)_{m\to n}=\biggl|\int_{-\infty}^{+\infty}dx\,\Psi_{n}^{*(0)}(x,t)\Psi_{m}(x,t)\biggr|^{2}.
\end{eqnarray}
Using Eqs. \eqref{eq:função de onda do oscilador} and \eqref{eq:Psi-0} one can find that ${\cal P}(t)_{m\to n}=0$ for odd values of $|n-m|$, and \cite{Kim-OC-1989,Kim-PRA-1989}
\begin{eqnarray}
\nonumber
{\cal P}(t)_{m\to n}=\frac{2^{m+n}\min(m,n)!^{2}\left\{ \sinh[r(t)]\right\} ^{|n-m|}}{m!n!\cosh[r(t)]}\\
\label{eq:Prob-m-n-r}
\times\left[\sum_{k=\frac{|n-m|}{2}}^{\frac{n+m}{2}}\frac{\left(\begin{array}{c}
		\frac{n+m}{2}\\
		k
	\end{array}\right)\left(\begin{array}{c}
		\frac{n+m+2k-2}{4}\\
		\frac{n+m}{2}
	\end{array}\right)k!}{\left(k-\frac{|n-m|}{2}\right)!\cosh^{k}[r(t)]}\right]^{2},
\end{eqnarray}
for even values of $|n-m|$, where $\min(m,n)$ is the smallest value between $ m $ and $ n $.
Note that, the fact that Eq. \eqref{eq:Prob-m-n-r} is non-zero only for $|n-m|$ even is related to the parity of the harmonic potential \cite{Popov-SJETP-1969}. 
From Eqs. \eqref{eq:<N>-2} (making $n=0$) and \eqref{eq:Prob-m-n-r}, we can relate the probability $	{\cal P}(t)_{m\to n} $ to the mean number of excitations in the fundamental state $ \langle\hat{N}\rangle(0,t)$:
\begin{eqnarray}
\nonumber
{\cal P}(t)_{m\to n}=\frac{2^{m+n}\min(m,n)!^{2}\bigl[\langle\hat{N}\rangle(0,t)\bigr]^{\frac{|n-m|}{2}}}{m!n!\bigl[\langle\hat{N}\rangle(0,t)+1\bigr]^{\frac{1}{2}}}\\
\label{eq:Prob-m-n-N0}
	\times\left[\sum_{k=\frac{|n-m|}{2}}^{\frac{n+m}{2}}\frac{\left(\begin{array}{c}
			\frac{n+m}{2}\\
			k
		\end{array}\right)\left(\begin{array}{c}
			\frac{n+m+2k-2}{4}\\
			\frac{n+m}{2}
		\end{array}\right)k!}{\left(k-\frac{|n-m|}{2}\right)!\bigl[\langle\hat{N}\rangle(0,t)+1\bigr]^{\frac{k}{2}}}\right]^{2}.
\end{eqnarray}
It follows that if the mean number of excitations in the fundamental state is non-zero, then the oscillator will have non-zero probabilities of making transitions between different energy levels.
When we consider $m=n=0$ in Eq. \eqref{eq:Prob-m-n-r} [or in Eq. \eqref{eq:Prob-m-n-N0}], one has the probability of persistence in the fundamental state, and, as a consequence, one can also obtain the probability of excitation, given by $1-{\cal P}(t)_{0\,\to\,0}$ \cite{Tibaduiza-BJP-2020}.

\section{Oscillator with two frequency jumps}
\label{sec:model}

Now, we apply the formulas shown in Sec. \ref{sec:review} to 
investigate the model discussed in Ref. \cite{Tibaduiza-BJP-2020}, namely an oscillator with
\begin{eqnarray}
\label{eq:Tibaduiza}
\omega(t)=\begin{cases}
\omega_{0}, & t<0,\\
\omega_{1}, & 0<t<\tau,\\
\omega_{0}, & t>\tau,
\end{cases}
\end{eqnarray}
in which $\omega_{0}$ and $\omega_{1}$ are constant frequencies, and $ \tau $ is the length of the time interval between the frequency jumps.


\subsection{Solution and General Behavior of the $ \rho(t) $ Parameter}

Due to the form of Eq. \eqref{eq:Tibaduiza}, the $\rho(t)$ parameter can be written as
\begin{eqnarray}
	\label{eq:rho}
	\rho(t)=\begin{cases}
		\rho_{0}, & t<0,\\
		\rho_{1}(t), & 0<t<\tau,\\
		\rho_{2}(t), & t>\tau,
	\end{cases}
\end{eqnarray}
where $\rho_0$ is given in Eq. \eqref{eq:rho0}, and $\rho_{1}(t)$ and $\rho_{2}(t)$ are calculated next. 
%

\subsubsection{Interval $ 0<t<\tau $}
\label{sec:0-tau}

For the interval $0< t < \tau$, the equation to be solved is
\begin{eqnarray}
\label{eq:EP-rho1-Tibaduiza}
\ddot{\rho}_{1}(t)+\omega_{1}^{2}\rho_{1}(t)=\frac{1}{m_{0}^{2}\rho_{1}(t)^{3}},
\end{eqnarray}
with the conditions \cite{Ciftja-JPA-1999}
\begin{eqnarray}
\label{eq:cond-cont-rho1-Tibaduiza}
\rho_{1}\left(t=0\right)=\frac{1}{\sqrt{m_{0}\omega_{0}}},\;\;\dot{\rho}_{1}\left(t=0\right)=0.
\end{eqnarray}
The general solution for $\rho_{1}(t)$ is of the form \cite{Pinney-PAMS-1950,Lima-JMO-2009}
\begin{eqnarray}
\nonumber
\rho_{1}(t)=\bigl[A_{1}\sin^{2}\left(\omega_{1}t\right)+2C_{1}\sin\left(\omega_{1}t\right)\cos\left(\omega_{1}t\right)\\
\label{eq:rho1-geral-Tibaduiza}
+B_{1}\cos^{2}\left(\omega_{1}t\right)\bigr]^{\frac{1}{2}},
\end{eqnarray}
and the relationship between the constants $A_1$, $B_1$ and $C_1$ is 
\begin{eqnarray}
\label{eq:rel-A1-B1-C1-rho1-Tibaduiza}	
A_{1}B_{1}-C_{1}^{2}=\frac{1}{m_{0}^{2}\omega_{1}^{2}}.
\end{eqnarray}
Then, applying conditions \eqref{eq:cond-cont-rho1-Tibaduiza} to Eq. \eqref{eq:rho1-geral-Tibaduiza} and using relation \eqref{eq:rel-A1-B1-C1-rho1-Tibaduiza}, we get
\begin{eqnarray}
\label{eq:rho1-Tibaduiza}	
\rho_{1}(t)=\left[\frac{\omega_{0}\sin^{2}\left(\omega_{1}t\right)}{m_{0}\omega_{1}^{2}}+\frac{\cos^{2}\left(\omega_{1}t\right)}{m_{0}\omega_{0}}\right]^{\frac{1}{2}}.
\end{eqnarray}
%

\subsubsection{Interval $ t>\tau $}
\label{sec:t-tau}

In the interval $t>\tau$, the Ermakov-Pinney equation has the form
\begin{eqnarray}
\label{eq:EP-rho2-Tibaduiza}
\ddot{\rho}_{2}(t)+\omega_{0}^{2}\rho_{2}(t)=\frac{1}{m_{0}^{2}\rho_{2}(t)^{3}}.
\end{eqnarray}
The general solution of Eq. \eqref{eq:EP-rho2-Tibaduiza} is
\begin{eqnarray}
	\nonumber
	\rho_{2}(t)=\bigl[A_{2}\sin^{2}\left(\omega_{0}t\right)+2C_{2}\sin\left(\omega_{0}t\right)\cos\left(\omega_{0}t\right)\\
	\label{eq:rho2-Tibaduiza}
	+B_{2}\cos^{2}\left(\omega_{0}t\right)\bigr]^{\frac{1}{2}},
\end{eqnarray}
with the constants $A_2$, $B_2$ and $C_2$ determined from the relationship
\begin{eqnarray}
	\label{eq:rel-A2-B2-C2-rho2-Tibaduiza}	
	A_{2}B_{2}-C_{2}^{2}=\frac{1}{m_{0}^{2}\omega_{0}^{2}},
\end{eqnarray}
and the conditions for continuity 
\begin{eqnarray}
\label{eq:cond-cont-rho2-Tibaduiza}	
\rho_{1}(t=\tau)=\rho_{2}(t=\tau),\;\;\;\dot{\rho}_{1}(t=\tau)=\dot{\rho}_{2}(t=\tau).
\end{eqnarray}
Using Eqs. \eqref{eq:rho1-Tibaduiza}, \eqref{eq:rho2-Tibaduiza},  \eqref{eq:rel-A2-B2-C2-rho2-Tibaduiza} and \eqref{eq:cond-cont-rho2-Tibaduiza}, the results in
\begin{eqnarray}
\nonumber
A_{2}=\frac{1}{m_{0}\omega_{0}^{3}\omega_{1}^{2}}\Bigl\{\omega_{0}^{2}\omega_{1}^{2}+\bigl[(\omega_{0}^{4}-\omega_{1}^{4})\sin^{2}(\omega_{0}\tau)-\omega_{0}^{2}\omega_{1}^{2}\\
\nonumber
+\omega_{1}^{4}\bigr]\sin^{2}(\omega_{1}\tau)+2\omega_{0}\omega_{1}(\omega_{0}-\omega_{1})(\omega_{0}+\omega_{1})\sin(\omega_{0}\tau)\\
\nonumber
\times\cos(\omega_{0}\tau)\sin(\omega_{1}\tau)\cos(\omega_{1}\tau)\Bigr\},\\
\label{eq:A2-Tibaduiza}\\
\nonumber
B_{2}=\frac{1}{m_{0}\omega_{0}^{3}\omega_{1}^{2}}\Bigl\{\omega_{0}^{2}\omega_{1}^{2}+\bigl[\left(\omega_{1}^{4}-\omega_{0}^{4}\right)\sin^{2}\left(\omega_{0}\tau\right)-\omega_{0}^{2}\omega_{1}^{2}\\
\nonumber
+\omega_{0}^{4}\bigr]\sin^{2}\left(\omega_{1}\tau\right)-2\omega_{0}\omega_{1}\left(\omega_{0}-\omega_{1}\right)\left(\omega_{0}+\omega_{1}\right)\sin\left(\omega_{0}\tau\right)\\
\nonumber
\times\cos\left(\omega_{0}\tau\right)\sin\left(\omega_{1}\tau\right)\cos\left(\omega_{1}\tau\right)\Bigr\},\\
\label{eq:B2-Tibaduiza}\\
\nonumber
C_{2}=\frac{1}{m_{0}\omega_{0}^{3}\omega_{1}^{2}}\Bigl\{\bigl[\left(\omega_{0}^{2}+\omega_{1}^{2}\right)\sin\left(\omega_{0}\tau\right)\cos\left(\omega_{0}\tau\right)\sin\left(\omega_{1}\tau\right)\\
\nonumber
-2\omega_{0}\omega_{1}\left(\sin^{2}\left(\omega_{0}\tau\right)-1/2\right)\cos\left(\omega_{1}\tau\right)\bigr]\left(\omega_{0}-\omega_{1}\right)\\
\nonumber
\times\left(\omega_{0}+\omega_{1}\right)\sin\left(\omega_{1}\tau\right)\Bigr\}.\\
\label{eq:C2-Tibaduiza}
\end{eqnarray}
%

\subsubsection{General Behavior}

\begin{figure}[h]
	\centering
	\epsfig{file=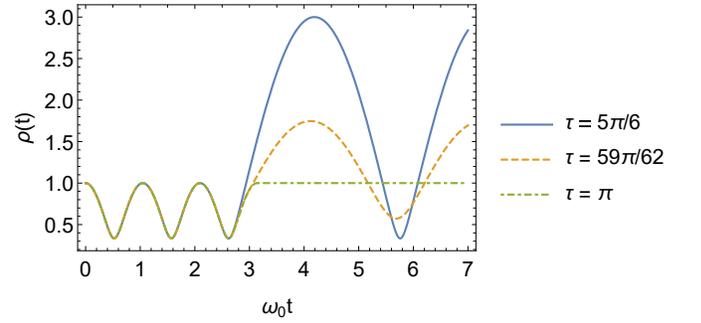,  width=1.0 \linewidth}  
	\caption{General behavior of $\rho(t)$ as a function of $ \omega_{0}t $, with  $\omega_1=3\omega_0$ and different values of $ \tau $ (we consider, for simplicity, $ m_0=\omega_0=1 $ and $ \hbar=1 $ in arbitrary units). For $ \tau=[\tau_{l+1/2}]_{l=2}=5\pi/6 $, we note that the amplitude of oscillation for the $ \rho(t) $ parameter is maximum. For $ \tau=59\pi/62 $, the amplitude of oscillation is intermediate. Finally, when $ \tau=[\tau_{l}]_{l=3}=\pi $, there is no oscillation at all, and it follows that $\rho(t)$ becomes time independent.}
	\label{fig:rho-Tibaduiza}
\end{figure}

The general solution for the $\rho(t)$ parameter
is given by Eq. \eqref{eq:rho}, with $ \rho_{0} $, $\rho_1(t)$ and $\rho_2(t)$ given by
Eqs. \eqref{eq:rho0}, \eqref{eq:rho1-Tibaduiza}, \eqref{eq:rho2-Tibaduiza}, \eqref{eq:A2-Tibaduiza}, \eqref{eq:B2-Tibaduiza},
and \eqref{eq:C2-Tibaduiza}.
From these equations, it can be seen that the $ \rho(t) $ parameter is a periodic function of time. 
Moreover, even when the frequency returns to its initial value $ \omega_{0} $, this parameter will still, in general, be a periodic function of time.
However, if we define $\tau_{u}=u\pi/\omega_{1}$ ($u>0$), and make $\tau=\tau_{l}$, where $l\in\mathbb{N}$, the $ \rho(t) $ parameter returns to $ \rho_{0} $ [Eq. \eqref{eq:rho0}],
which means that although the oscillator feels the effect of the change in its frequency when it jumps from $\omega_{0}$ to $\omega_{1}$, if the frequency returns to $\omega_{0}$ at $\tau=\tau_{l}$, for $t> \tau_l$ the oscillator behaves as if nothing happened.
In other words, if $\tau=\tau_l$, the abrupt change in the frequency is imperceptible to the oscillator when $t> \tau_l$.
On the other hand, when $ \tau=\tau_{l+1/2}$, the $ \rho(t) $ parameter reaches its maximum value.
The behavior of $ \rho(t) $ is shown in Fig. \ref{fig:rho-Tibaduiza}.

%

\subsection{Squeeze Parameters}

Because the frequency varies abruptly, squeezing occurs in the system \cite{Tibaduiza-BJP-2020,Janszky-PRA-1992,Kiss-PRA-1994}. 
Thus, now, we calculate the parameters $ r(t) $ and $\phi(t)$ associated with the model 
in Eq. \eqref{eq:Tibaduiza}.
We show that our results for these parameters agree with those found in Ref. \cite{Tibaduiza-BJP-2020}
via an exact algebraic method.

\subsubsection{Parameter $ r(t) $}

\begin{figure}[h]
	\centering
	\epsfig{file=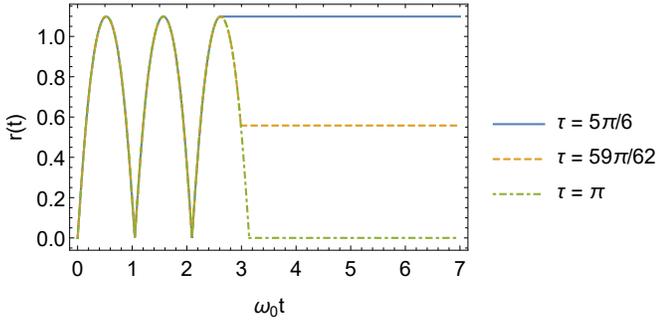,  width=1.0 \linewidth}  
	\caption{Behavior of the squeeze parameter $ r(t) $ as a function of $ \omega_{0}t $, where $ \omega_{1}=3\omega_{0} $ (we consider $\omega_0=1 $ in arbitrary units). }
	\label{fig:r-tibaduiza}
\end{figure}

The parameter $ r(t)  $ for any time interval is given by
\begin{eqnarray}
\label{eq:r-geral}
r(t)=\begin{cases}
	0, & t<0,\\
	r_{1}(t), & 0<t<\tau,\\
	r_{2}(t), & t>\tau.
\end{cases}
\end{eqnarray}
Note that $r(t<0)=0$ because the frequency of the oscillator is time independent in this interval. 
Using Eq. \eqref{eq:rho1-Tibaduiza} in Eq. \eqref{eq:r} we obtain, for the interval $ 0<t<\tau $, the squeezing parameter $r_1(t)$, where
\begin{eqnarray}
\label{eq:r1-tibaduiza}
r_{1}(t)=\cosh^{-1}\Biggl\{\sqrt{1+\biggl(\frac{\omega_{1}^{2}-\omega_{0}^{2}}{2\omega_{0}\omega_{1}}\biggr)^{2}\sin^{2}(\omega_{1}t)}\Biggr\},
\end{eqnarray}
which is a result that agrees with the one found in Ref. \cite{Tibaduiza-BJP-2020}. 
For the interval $ t>\tau $, using Eq. \eqref{eq:rho2-Tibaduiza} in Eq. \eqref{eq:r}, we find that 
$r_2(t)$, with
\begin{eqnarray}
\label{eq:r2-tibaduiza}
r_2(t)=r_1(\tau),
\end{eqnarray}
that also agrees with Ref. \cite{Tibaduiza-BJP-2020}.
Note that for $\tau=\tau_l$, one has $r_{2}(t)=0$. In Fig. \ref{fig:r-tibaduiza} 
(also found in Ref. \cite{Tibaduiza-BJP-2020}), one can see
the behavior of $r(t)$ for some values of $\tau$.
%

\subsubsection{Parameter $ \phi(t) $}

The squeeze phase $ \phi(t) $ for any time interval has the form  
\begin{eqnarray}
	\label{eq:phi-geral}
\phi(t)=\begin{cases}
	\text{undefined}, & t<0,\\
	\phi_{1}(t), & 0<t<\tau,\\
	\phi_{2}(t), & t>\tau,
\end{cases}
\end{eqnarray}
where $ \phi(t) $ for $ t< 0 $ is undefined because there is no squeeze in this interval. 
Using Eqs. \eqref{eq:rho1-Tibaduiza} and \eqref{eq:r1-tibaduiza} in Eq. \eqref{eq:phi}, we obtain that the squeeze phase for the interval $ 0<t<\tau $, $ \phi_{1}(t) $, is given by
\begin{eqnarray}
\nonumber
\phi_{1}(t)=\cos^{-1}\Bigl\{\left(\omega_{0}^{4}-\omega_{1}^{4}\right)\sin^{2}(\omega_{1}t)\bigl[\bigl(4\omega_{0}^{2}\omega_{1}^{2}+\bigl(\omega_{1}^{2}\\
\label{eq:phi-1-tibaduiza}
-\omega_{0}^{2}\bigr)^{2}\sin^{2}(\omega_{1}t)\bigr)\left(\omega_{1}^{2}-\omega_{0}^{2}\right)^{2}\sin^{2}(\omega_{1}t)\bigr]^{-\frac{1}{2}}\Bigr\},
\end{eqnarray}
which also agrees with Ref. \cite{Tibaduiza-BJP-2020}. 
To calculate the squeeze phase in the interval $ t>\tau $, the reasoning is analogous, simply substituting Eqs. \eqref{eq:rho2-Tibaduiza} and \eqref{eq:r2-tibaduiza} into Eq. \eqref{eq:phi}.
\begin{figure}[h]
	\centering
	\epsfig{file=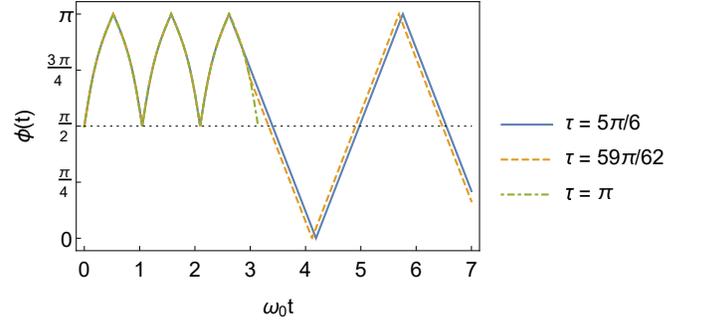,  width=1.0 \linewidth}  
	\caption{Behavior of the squeeze phase $ \phi(t) $ as a function of $ \omega_{0}t $, where $ \omega_{1}=3\omega_{0} $ (we consider $\omega_0=1 $ in arbitrary units). }
	\label{fig:phi-tibaduiza}
\end{figure}

From Fig. \ref{fig:phi-tibaduiza}, it can be seen that the squeezing phase will continue to vary in time even for $t>\tau$.
Due of this time dependence, the fluctuations of the $ \hat{x} $ and $ \hat{p} $ operators will continue to depend on time in this interval, as we will see later in Sec. \ref{sec:quantum fluctuations} [see Eqs. \eqref{eq:Delta x^2} and \eqref{eq:Delta p^2}].
Another point to be observed, in Fig. \ref{fig:phi-tibaduiza}, concerns the behavior of the squeezing phase in the interval $ t>\tau $, when $ \tau=\tau_l $ (in the specific case of Fig. \ref{fig:phi-tibaduiza}, $\tau_l=\pi$). Since the squeeze parameter $ r_2(t) $ [Eq. \eqref{eq:r2-tibaduiza}] is zero in this case, the system is no longer squeezed. Consequently, the squeeze phase is undefined for $ \tau_l $. Therefore, its effect on the system will be negligible because there will be no more squeezing.

\subsection{Quantum Fluctuations}\label{sec:quantum fluctuations}

\begin{figure}[h]
	\centering
	\epsfig{file=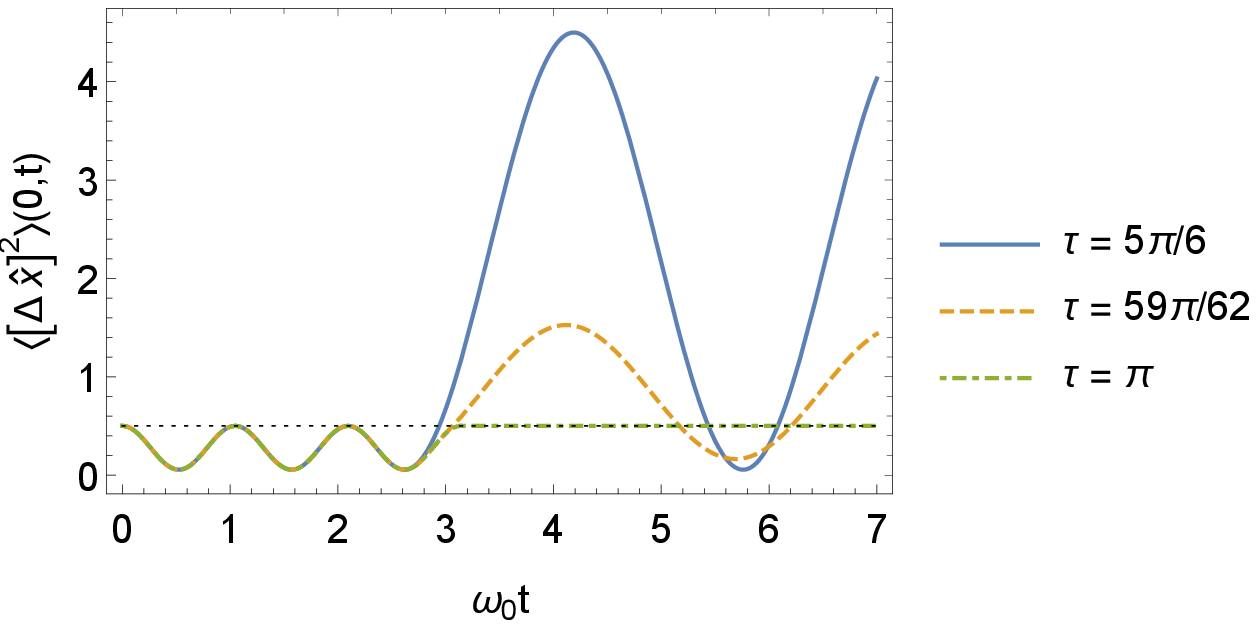,  width=1.0 \linewidth}  
	\caption{Behavior of the $ \langle[\Delta\hat{x}]^{2}\rangle(n,t) $ for $ n=0 $ as a function of $ \omega_{0}t $, where $ \omega_{1}=3\omega_{0} $ (we consider, for simplicity, $ m_0=\omega_0=1 $ and $ \hbar=1 $ in arbitrary units). }
	\label{fig:delta-x-tibaduiza}
\end{figure}

The variance of the $ \hat{x} $ operator for any time interval is given by
\begin{eqnarray}
	\label{eq:delta-x-geral}
\langle[\Delta\hat{x}]^{2}\rangle(n,t)=\begin{cases}
	\langle[\Delta\hat{x}]_{0}^{2}\rangle(n), & t<0,\\
	\langle[\Delta\hat{x}]_{1}^{2}\rangle(n,t), & 0<t<\tau,\\
	\langle[\Delta\hat{x}]_{2}^{2}\rangle(n,t), & t>\tau,
\end{cases}
\end{eqnarray}
where \cite{Sakurai-Quantum-Mechanics-2021}
\begin{eqnarray}
\langle[\Delta\hat{x}]_{0}^{2}\rangle(n)=\left(n+\frac{1}{2}\right)\frac{\hbar}{m_{0}\omega_{0}}.
\end{eqnarray}
Substituting Eqs. \eqref{eq:r1-tibaduiza} and \eqref{eq:phi-1-tibaduiza} into \eqref{eq:Delta x^2} we find, for the $ \hat{x} $ operator in the interval $ 0<t<\tau $,
\begin{eqnarray}
	\nonumber
\langle[\Delta\hat{x}]_{1}^{2}\rangle(n,t)=\left[\frac{\omega_{0}^{2}}{\omega_{1}^{2}}\sin^{2}(\omega_{1}t)+\cos^{2}(\omega_{1}t)\right]\langle[\Delta\hat{x}]_{0}^{2}\rangle(n),\\
\label{eq:Delta x1^2-tibaduiza}
\end{eqnarray}
which is in agreement with Refs. \cite{Tibaduiza-BJP-2020,Janszky-PRA-1992}. For the interval $ t>\tau $, the procedure is analogous, using Eqs. \eqref{eq:r}, \eqref{eq:phi} and \eqref{eq:rho2-Tibaduiza} in Eq. \eqref{eq:Delta x^2}. In Fig. \ref{fig:delta-x-tibaduiza}, we show the behavior of $ \langle[\Delta\hat{x}]^{2}\rangle(n,t) $ for $ n=0 $. 
%

\begin{figure}[h]
	\centering
	\epsfig{file=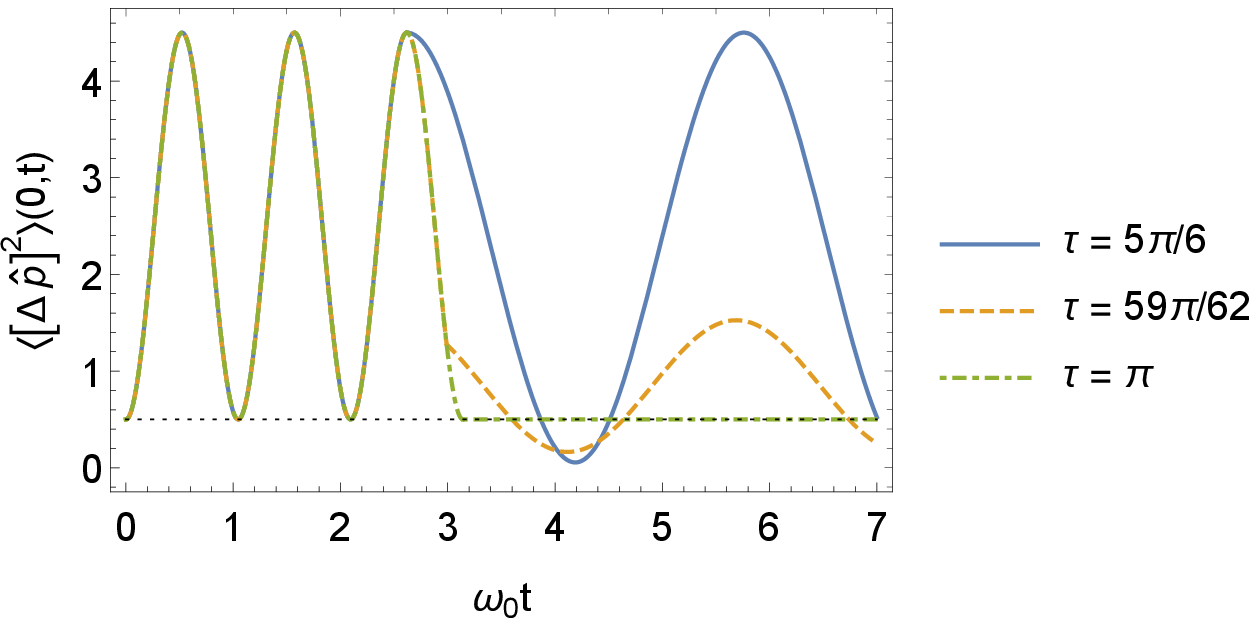,  width=1.0 \linewidth}  
	\caption{Behavior of the $ \langle[\Delta\hat{p}]^{2}\rangle(n,t) $ for $ n=0 $ as a function of $ \omega_{0}t $, where $ \omega_{1}=3\omega_{0} $ (we consider, for simplicity, $ m_0=\omega_0=1 $ and $ \hbar=1 $ in arbitrary units). }
	\label{fig:delta-p-tibaduiza}
\end{figure}

Similarly, for the variance of the $ \hat{p} $ operator, we have
\begin{eqnarray}
	\label{eq:delta-p-geral}
\langle[\Delta\hat{p}]^{2}\rangle(n,t)=\begin{cases}
	\langle[\Delta\hat{p}]_{0}^{2}\rangle(n), & t<0,\\
	\langle[\Delta\hat{p}]_{1}^{2}\rangle(n,t), & 0<t<\tau,\\
	\langle[\Delta\hat{p}]_{2}^{2}\rangle(n,t), & t>\tau,
\end{cases}
\end{eqnarray}
being \cite{Sakurai-Quantum-Mechanics-2021}
\begin{eqnarray}
\langle[\Delta\hat{p}]_{0}^{2}\rangle(n)=\left(n+\frac{1}{2}\right)\hbar m_{0}\omega_{0}.
\end{eqnarray}
Through Eqs. \eqref{eq:r1-tibaduiza}, \eqref{eq:phi-1-tibaduiza} and \eqref{eq:Delta p^2}, we find, for the interval $ 0<t<\tau $, the expression \cite{Tibaduiza-BJP-2020,Janszky-PRA-1992}
\begin{eqnarray}
	\nonumber
\langle[\Delta\hat{p}]_{1}^{2}\rangle(n,t)=\left[\frac{\omega_{1}^{2}}{\omega_{0}^{2}}\sin^{2}(\omega_{1}t)+\cos^{2}(\omega_{1}t)\right]\langle[\Delta\hat{p}]_{0}^{2}\rangle(n).\\
\label{eq:Delta p1^2-tibaduiza}
\end{eqnarray}
To calculate the variance of the operator $ \hat{p} $ in the interval $ t>\tau $, we must use Eqs. \eqref{eq:r}, \eqref{eq:phi} and \eqref{eq:rho2-Tibaduiza} in Eq. \eqref{eq:Delta p^2}. The general behavior of $ \langle[\Delta\hat{p}]^{2}\rangle(n,t) $, when $n=0$, is schematized in Fig. \ref{fig:delta-p-tibaduiza}.

\begin{figure}[h]
\centering
\epsfig{file=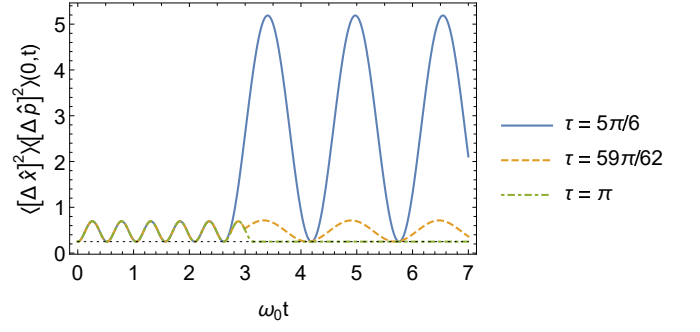,  width=1.0 \linewidth}  
\caption{Behavior of the $ \langle[\Delta\hat{x}]^{2}\rangle(n,t)\langle[\Delta\hat{p}]^{2}\rangle(n,t) $ for $ n=0 $ as a function of $ \omega_{0}t $, where $ \omega_{1}=3\omega_{0} $ (we consider, for simplicity, $ m_0=\omega_0=1 $ and $ \hbar=1 $ in arbitrary units). }
\label{fig:incerteza-tibaduiza}
\end{figure}

Thus, it is direct to see that the uncertainty relation between these operators in the interval $ 0<t<\tau $ has the form
\begin{eqnarray}
\nonumber
\langle[\Delta\hat{x}]_{1}^{2}\rangle(n,t)\langle[\Delta\hat{p}]_{1}^{2}\rangle(n,t)\geq\Biggl\{1+\biggl[\biggl(\frac{\omega_{1}^{2}-\omega_{0}^{2}}{\omega_{0}\omega_{1}}\biggr)\\
\times\sin(\omega_{1}t)\cos(\omega_{1}t)\biggr]^{2}\Biggr\}\langle[\Delta\hat{x}]_{0}^{2}\rangle(n)\langle[\Delta\hat{p}]_{0}^{2}\rangle(n),
\label{eq:inc-1}
\end{eqnarray}
and the uncertainty relation for the interval $ t>\tau $ is obtained in a similar way. 
Clearly, when $ \omega_{1}=\omega_{0} $, the uncertainty relation [Eq. \eqref{eq:inc-1}] falls back to the uncertainty relation of a time independent oscillator \cite{Sakurai-Quantum-Mechanics-2021}.
Furthermore, the uncertainty relation for an oscillator with time independent frequency is also reobtained when $\tau=\tau_l$, as shown in Fig. \ref{fig:incerteza-tibaduiza}. 

The results found by us (via the LR method) given in Eqs.
\eqref{eq:r1-tibaduiza}, \eqref{eq:r2-tibaduiza}, \eqref{eq:phi-1-tibaduiza}, \eqref{eq:Delta x1^2-tibaduiza}, and \eqref{eq:Delta p1^2-tibaduiza} are in agreement with those found in 
Ref. \cite{Tibaduiza-BJP-2020}.
Hereafter, we use the LR method to obtain new results concerning the model given in Eq. \eqref{eq:Tibaduiza}.

\subsection{Mean Energy}

The expected value of the Hamiltonian operator is identified as the mean energy of the system, that is, $E\left(n,t\right)=\langle\hat{H}(t)\rangle(n,t) $.
The mean energy for the model in Eq. \eqref{eq:Tibaduiza} can be written as
\begin{eqnarray}
	E(n,t)=\begin{cases}
		E_{0}(n), & t<0,\\
		E_{1}(n,t), & 0<t<\tau,\\
		E_{2}(n,t), & t>\tau,
	\end{cases}
\end{eqnarray}
wherein \cite{Sakurai-Quantum-Mechanics-2021}
\begin{eqnarray}
	\label{eq:E0}
	E_{0}(n)=\left(n+\frac{1}{2}\right)\hbar\omega_{0}.
\end{eqnarray}

For the interval $ 0<t<\tau $, through Eqs. \eqref{eq:Delta x1^2-tibaduiza}, \eqref{eq:Delta p1^2-tibaduiza}, \eqref{eq:E0} and \eqref{eq:<H>}, we find that the mean energy is time independent, given by
\begin{eqnarray}
	\label{eq:<H>-1-tibaduiza}	
	E_{1}(n,t)=\frac{1}{2}\left(1+\frac{\omega_{1}^{2}}{\omega_{0}^{2}}\right)E_{0}(n).
\end{eqnarray}
When $ \omega_{1}=\omega_{0} $, Eq. \eqref{eq:<H>-1-tibaduiza} reduces to $ E_{1}(n,t)=E_{0}(n) $, as expected. 
%
%
Note that for $\omega_1/\omega_0<1$, $E_{1}(n,t)<E_{0}(n)$, whereas for $\omega_1/\omega_0>1$, we
have $E_{1}(n,t)>E_{0}(n)$.
When $\omega_{1}=0$, which means that the system is free in interval $ 0<t<\tau $, we have $E_{1}(n,t)=E_{0}(n)/2$. 
This can also be obtained by making $\omega_{0}\gg\omega_{1} $, which leads to $E_{1}(n,t)\approx E_{0}(n)/2$.

\begin{figure}[h]
	\centering  
	\epsfig{file=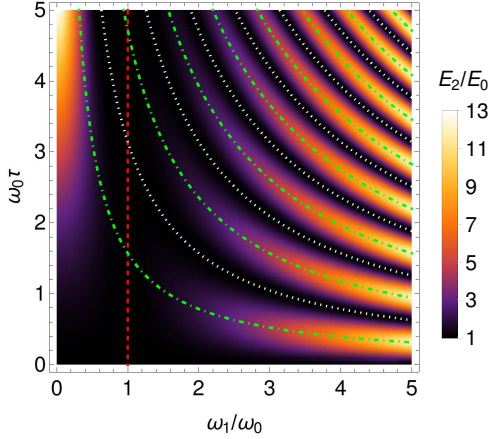,  width=0.75 \linewidth} 
	\caption{Ratio $ E_{2}(n,t)/E_{0}(n) $ as a function of $ \omega_{0}\tau $ and $ \omega_{1}/\omega_0 $. The dashed line corresponds to $ \omega_{1}=\omega_{0}$. The dotted lines correspond to $\tau=\tau_l$. The dot-dashed lines correspond to $ \tau=\tau_{l+1/2}$.}
	\label{fig:energia-E2-E0-tibaduiza}
\end{figure}

For the interval $ t>\tau $, from Eqs. \eqref{eq:r}, \eqref{eq:phi}, \eqref{eq:<H>}, \eqref{eq:rho2-Tibaduiza}, and \eqref{eq:E0}, we have that the mean energy $E_{2}(n,t)$ is given by
\begin{eqnarray}
	E_{2}(n,t)=\biggl[1+\frac{1}{2}\biggl(\frac{\omega_{1}^{2}-\omega_{0}^{2}}{\omega_{0}\omega_{1}}\biggr)^{2}\sin^{2}(\omega_{1}\tau)\biggr]E_{0}(n).
	\label{eq:<H>-2-tibaduiza}
\end{eqnarray}
Note that Eq. \eqref{eq:<H>-2-tibaduiza} is independent of $t$, and 
$E_{2}(n,t)\geq E_{0}(n)$, even for $ \omega_{1}<\omega_{0}$.
The behavior of the ratio $ E_{2}(n,t)/E_{0}(n) $ is illustrated in Fig. \ref{fig:energia-E2-E0-tibaduiza}. 
For $ \omega_{1}=\omega_{0}$ (dashed line in Fig. \ref{fig:energia-E2-E0-tibaduiza}), Eq. \eqref{eq:<H>-2-tibaduiza} recovers $E_{2}(n,t)=E_{0}(n)$, as expected.
Furthermore, for $\tau=\tau_l$
(dotted lines in Fig. \ref{fig:energia-E2-E0-tibaduiza}), Eq. \eqref{eq:<H>-2-tibaduiza} also gives $ E_{2}(n,t)=E_{0}(n)$.
In particular, the energy of this system is maximized when $ \tau=\tau_{l+1/2}$
(dot-dashed lines in Fig. \ref{fig:energia-E2-E0-tibaduiza}).
For $ \omega_{1}/\omega_{0}\to 0 $, we obtain $ E_{2}(n,t)=\left(1+\omega_{0}^{2}\tau^{2}/2\right)E_{0}(n)$.
Moreover, from Eqs. \eqref{eq:r2-tibaduiza} and \eqref{eq:<H>-2-tibaduiza}, we obtain
\begin{eqnarray}
	\label{eq:E_2-r_2}
	E_{2}(n,t)=\left\{ 2\cosh^{2}\left[r_{2}(t)\right]-1\right\} E_{0}(n).
\end{eqnarray}
Thus, while there is squeeze, $E_{2}(n,t)>E_{0}(n)$.
Therefore, the squeezing caused by the frequency jumps results in an increase in the mean energy of the oscillator, with respect to its initial energy $ E_0(n) $.

\begin{figure}[h]
	\centering
	\epsfig{file=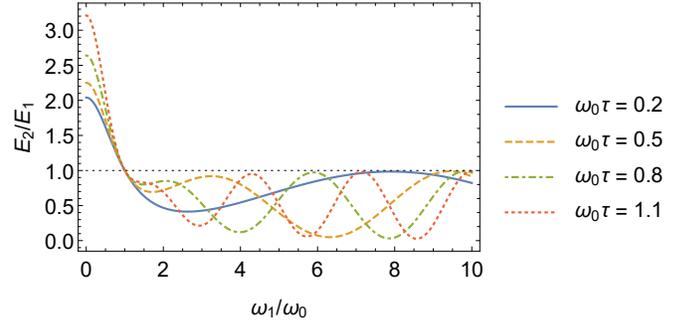,  width=1.0 \linewidth}  
	\caption{Ratio $ E_{2}(n,t)/E_{1}(n,t) $ as a function of $ \omega_{1}/\omega_{0}$. For $  \omega_{1}/\omega_{0}<1 $, $ E_{2}(n,t)/E_{1}(n,t)>1 $, whereas for $ \omega_{1}/\omega_{0}>1 $, $ E_{2}(n,t)/E_{1}(n,t)\leq 1 $.}
	\label{fig:E2-E1-w1-w0}
\end{figure}

It is interesting to investigate the behavior of  $E_2(n,t)/E_1(n,t)$.
Unlike the ratio $ E_{2}(n,t)/E_{0}(n) $, which is such that 
$ E_{2}(n,t)/E_{0}(n)\geq 1$, the ratio $ E_{2}(n,t)/E_{1}(n,t) $ can be lesser than, equal to, or greater than one, as shown in Fig. \ref{fig:E2-E1-w1-w0}. 
More specifically, for $ \omega_{1}/\omega_{0}<1 $, we have $ E_{2}(n,t)/E_{1}(n,t)>1 $, and when $ \omega_{1}/\omega_{0}\to 0 $, we find 
$E_{2}(n,t)=\left(2+\omega_{0}^{2}\tau^{2}\right)E_{1}(n,t)$.
For $ \omega_{1}/\omega_{0}>1 $, the ratio $ E_{2}(n,t)/E_{1}(n,t) $ oscillates between zero and one (see Fig. \ref{fig:E2-E1-w1-w0}). 
We highlight that, besides the trivial case $ \omega_{1}=\omega_{0} $, there are other values of the ratio $ \omega_{1}/\omega_{0} $ that result in $ E_{2}(n,t)=E_{1}(n,t) $.

%

\subsection{Mean Number of Excitations}

The mean number of excitations, $\langle\hat{N}\rangle(n,t)$, for the model in Eq. \eqref{eq:Tibaduiza},
is given by
\begin{eqnarray}
	\label{eq:<N>-geral}
\langle\hat{N}\rangle(n,t)=\begin{cases}
	\langle\hat{N}\rangle_{0}(n), & t<0,\\
	\langle\hat{N}\rangle_{1}(n,t), & 0<t<\tau,\\
	\langle\hat{N}\rangle_{2}(n,t), & t>\tau,
\end{cases}
\end{eqnarray}
where \cite{Sakurai-Quantum-Mechanics-2021}
\begin{eqnarray}
\langle\hat{N}\rangle_{0}(n)=n.
\end{eqnarray}
Given this, for the interval $ 0<t<\tau $, by means of Eqs. \eqref{eq:<N>-2} and \eqref{eq:r1-tibaduiza}, we have
\begin{eqnarray}
	\nonumber
\langle\hat{N}\rangle_{1}(n,t)=n+\left(n+\frac{1}{2}\right)\biggl[\frac{1}{2}\biggl(\frac{\omega_{1}^{2}-\omega_{0}^{2}}{\omega_{0}\omega_{1}}\biggr)^{2}\sin^{2}(\omega_{1}t)\biggr].\\
\label{eq:<N>-1-tibaduiza}
\end{eqnarray}
We remark the time dependence in $\langle\hat{N}\rangle_{1}(n,t)$, whereas the mean energy in the interval $ 0<t<\tau $ [Eq. \eqref{eq:<H>-1-tibaduiza}] is time independent.

For the interval $ t>\tau $, through Eqs. \eqref{eq:<N>-2} and \eqref{eq:r2-tibaduiza}, we obtain $ \langle\hat{N}\rangle_{2}(n,t) $, given by
\begin{eqnarray}	
\langle\hat{N}\rangle_{2}(n,t)=\langle\hat{N}\rangle_{1}(n,\tau).
\label{eq:<N>-2-tibaduiza}
\end{eqnarray}

\begin{figure}[h]
	\centering
	\epsfig{file=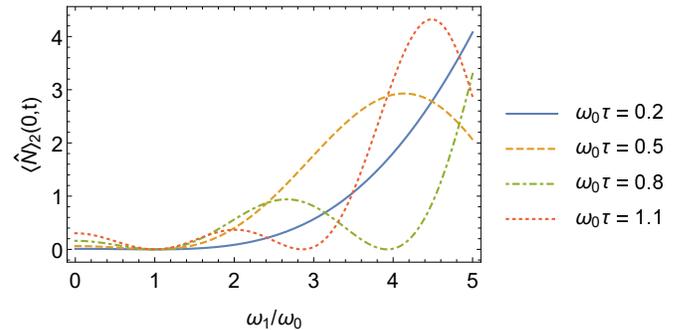, width=1.0 \linewidth}  
	\caption{Some examples of mean number of excitations $ \langle\hat{N}\rangle_{2}(0,t)$ that an oscillator could undergo as a function of $ \omega_{1}/\omega_{0} $ for different values of $ \omega_{0}\tau $.}
	\label{fig:N-tau-w1-w0}
\end{figure}

We also remark that for $ \tau=\tau_l $, the behavior of the system 
returns to that of the time independent oscillator found before the frequency jumps, i.e, $ \langle\hat{N}\rangle_{2}(n,t)=\langle\hat{N}\rangle_{0}(n) $.
We highlight that there is excitation even for $ n=0 $, 
which means that, under jumps in its frequency, 
a quantum oscillator initially in its ground state can become excited
(a classical oscillator in its ground state would remain in the same state).
In addition, excitation can also occur when $ \omega_{1}/\omega_{0}<1 $, as shown in Fig. \ref{fig:N-tau-w1-w0}. 

\begin{figure}[h]
	\centering
	\epsfig{file=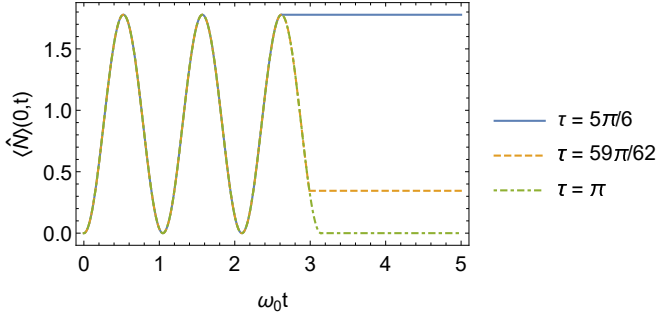, width=1.0 \linewidth}  
	\caption{Mean number of excitations $ \langle\hat{N}\rangle(0,t) $ as a function of $ \omega_{0}t $ for different values of $ \tau $, where $ \omega_{1}=3\omega_{0} $ (we consider $\omega_0=1 $ in arbitrary units).}
	\label{fig:N-t-tibaduiza}
\end{figure}

Using Eqs. \eqref{eq:<H>-2-tibaduiza} and \eqref{eq:<N>-2-tibaduiza}, we can also 
write $ E_{2}(n,t) $ as
\begin{eqnarray}
\label{eq:E_2-N_2}	
E_{2}(n,t)=\left[\langle\hat{N}\rangle_{2}(n,t)+\frac{1}{2}\right]\hbar\omega_{0},
\end{eqnarray}
which has the same structure as the expression for the energy eigenvalues of a time independent oscillator [see Eq. \eqref{eq:E0}]. 
The time evolution of $ \langle\hat{N}\rangle(0,t) $, given by Eq. \eqref{eq:<N>-geral} for $ n=0 $, is shown in Fig. \ref{fig:N-t-tibaduiza}.
%

\subsection{Transition Probability}

The general transition probability, $ {\cal P}(t)_{m\to n} $, for the model in Eq. \eqref{eq:Tibaduiza}, is given by
\begin{eqnarray}
{\cal P}(t)_{m\to n}=\begin{cases}
	\delta_{m,n}, & t<0,\\
	{\cal P}_{1}(t)_{m\to n}, & 0<t<\tau,\\
	{\cal P}_{2}(t)_{m\to n}, & t>\tau,
\end{cases}
\end{eqnarray}
with $ \delta_{m,n} $ being the Kronecker delta.
Using Eqs. \eqref{eq:Prob-m-n-r} and \eqref{eq:r1-tibaduiza} [or Eqs. \eqref{eq:Prob-m-n-N0} and \eqref{eq:<N>-1-tibaduiza} with $n=0$], we find, for the interval $ 0<t<\tau $, $ {\cal P}_{1}(t)_{m\to n} $, whose expression is 
%
\begin{eqnarray}
\nonumber
{\cal P}_{1}(t)_{m\to n}=\frac{2^{m+n}\min(m,n)!^{2}\Bigl[\Bigl(\frac{\omega_{1}^{2}-\omega_{0}^{2}}{2\omega_{0}\omega_{1}}\Bigr)\sin(\omega_{1}t)\Bigr]^{|n-m|}}{m!n!\Bigl[1+\Bigl(\frac{\omega_{1}^{2}-\omega_{0}^{2}}{2\omega_{0}\omega_{1}}\Bigr)^{2}\sin^{2}(\omega_{1}t)\Bigr]^{\frac{1}{2}}}\\
\nonumber
\times\left[\sum_{k=\frac{|n-m|}{2}}^{\frac{n+m}{2}}\frac{\left(\begin{array}{c}
		\frac{n+m}{2}\\
		k
	\end{array}\right)\left(\begin{array}{c}
		\frac{n+m+2k-2}{4}\\
		\frac{n+m}{2}
	\end{array}\right)k!}{\left(k-\frac{|n-m|}{2}\right)!\Bigl[1+\Bigl(\frac{\omega_{1}^{2}-\omega_{0}^{2}}{2\omega_{0}\omega_{1}}\Bigr)^{2}\sin^{2}(\omega_{1}t)\Bigr]^{\frac{k}{2}}}\right]^{2},\\
\label{eq:Prob-m-n-r-Tibaduiza-1}
\end{eqnarray}
for even values of $|n-m|$, and ${\cal P}(t)_{m\to n}=0$ for odd values of $|n-m|$.
%
Since the $ r_1(t) $ parameter [Eq. \eqref{eq:r1-tibaduiza}], in this interval, is explicitly time-dependent, consequently, the transition probability $ m\to n $ will also depend.

\begin{figure}[h]
	\centering
	\epsfig{file=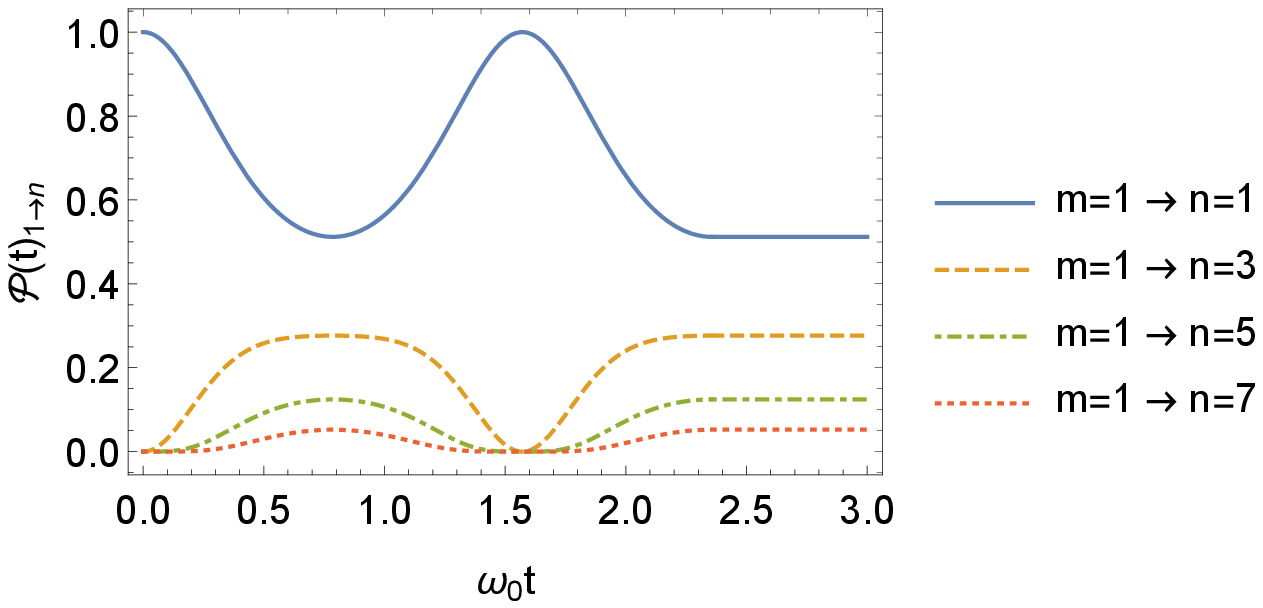,  width=1.0 \linewidth}  
	\caption{Behavior of $ {\cal P}(t)_{1\to n} $ as a function of $ \omega_{0}t $, with $ \omega_{1}=2\omega_{0} $ and $ \tau=3\tau_1/2=3\pi/4 $ (we consider $\omega_0=1 $ in arbitrary units).}
	\label{fig:P-m-n-tibaduiza}
\end{figure}

\begin{figure}[h]
	\centering
	\epsfig{file=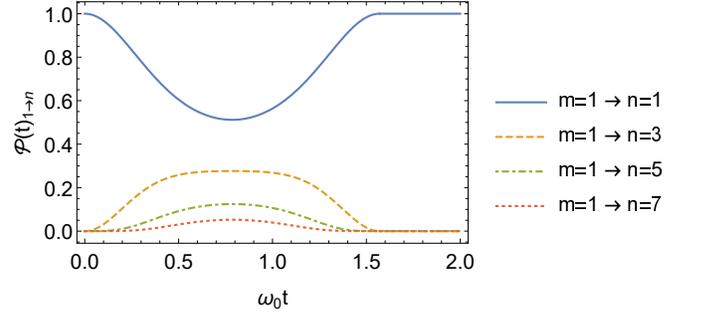,  width=1.0 \linewidth}  
	\caption{Behavior of $ {\cal P}(t)_{1\to n} $ as a function of $ \omega_{0}t $, with $ \omega_{1}=2\omega_{0} $ and $ \tau=\tau_1=\pi/2$ (we consider $\omega_0=1 $ in arbitrary units).}
	\label{fig:fig-P-1-n-tau-mathematica-tibaduiza}
\end{figure}

For the interval $ t>\tau $, using Eq. \eqref{eq:r2-tibaduiza} in Eq. \eqref{eq:Prob-m-n-r} [or Eqs. \eqref{eq:Prob-m-n-N0} and \eqref{eq:<N>-2-tibaduiza} with $n=0$], we see that the result is 
\begin{eqnarray}
{\cal P}_{2}(t)_{m\to n}={\cal P}_{1}(\tau)_{m\to n}.
\label{eq:Prob-m-n-r-Tibaduiza-2}
\end{eqnarray}
Thus, even when the frequency returns to $ \omega_{0} $ after an instant $ \tau $, one can find a non-zero $ m\to n $ transition probability, depending on the value of $ \tau $.
It is noticeable, from Eq. \eqref{eq:Prob-m-n-r-Tibaduiza-1}, that $ {\cal P}(t)_{m\to n} = {\cal P}(t)_{n\to m}$, with such symmetry being a consequence of the parity of the potential in Eq. \eqref{eq:hamiltoniano do oscilador} \cite{Popov-SJETP-1969}. 
The behavior of $ {\cal P}(t)_{1\to n} $ is illustrated in Figs. \ref{fig:P-m-n-tibaduiza} (for $\tau=3\tau_1/2$)
and \ref{fig:fig-P-1-n-tau-mathematica-tibaduiza} (for $\tau=\tau_1$). 
In Fig. \ref{fig:P-m-n-tibaduiza}, as $n$ increases, $ {\cal P}(t)_{1\to n} $ decreases. We highlight that for $t>\tau=\tau_1$ in Fig. \ref{fig:fig-P-1-n-tau-mathematica-tibaduiza}, ${\cal P}(t)_{1\to 1} = 1$ or, in other words, the oscillator remains in its same initial state.
In addition, for $m=0$, Eq. \eqref{eq:Prob-m-n-r-Tibaduiza-2} gives
\begin{eqnarray}
	\nonumber
	{\cal P}_{2}(t)_{0\,\to\,n}=\frac{n!\left(\frac{\omega_{1}^{2}-\omega_{0}^{2}}{2\omega_{0}\omega_{1}}\right)^{n}\sin^{n}(\omega_{1}\tau)}{2^{n}\left(\frac{n}{2}!\right)^{2}\left[1+\left(\frac{\omega_{1}^{2}-\omega_{0}^{2}}{2\omega_{0}\omega_{1}}\right)^{2}\sin^{2}(\omega_{1}\tau)\right]^{\frac{n+1}{2}}},\\
	\label{eq:P-0-n-tibaduiza}
\end{eqnarray}
where $ n=0,2,4,\ldots, $ recovering one of the results found in Ref. \cite{Tibaduiza-BJP-2020}.
Thus, Eq. \eqref{eq:Prob-m-n-r-Tibaduiza-2} generalizes the result for the transition probability found in Ref. \cite{Tibaduiza-BJP-2020}.
By also making $ n=0 $ in Eq. \eqref{eq:P-0-n-tibaduiza}, we find the probability of the oscillator of persisting in the fundamental state, and the probability of the oscillator being excited, after the frequency returns to $ \omega_0 $,
which is given by $1-{\cal P}_{2}(t)_{0\,\to\,0}$.
%

\section{Final Remarks}\label{sec:final}

Using the Lewis-Riesenfeld method, we investigated the dynamics of a quantum harmonic oscillator that undergoes two abrupt jumps in its frequency [Eq. \eqref{eq:Tibaduiza}].
We reobtained the analytical formulas of Ref. \cite{Tibaduiza-BJP-2020} 
for the squeeze parameters [Eqs. \eqref{eq:r1-tibaduiza}, \eqref{eq:r2-tibaduiza}, and \eqref{eq:phi-1-tibaduiza}], the quantum fluctuations of the position [Eq. \eqref{eq:Delta x1^2-tibaduiza}] and momentum [Eq. \eqref{eq:Delta p1^2-tibaduiza}] operators, and the probability amplitude of a transition from the fundamental state to an arbitrary energy eigenstate [Eq. \eqref{eq:P-0-n-tibaduiza}].
We also obtained expressions for the mean energy value [Eqs. \eqref{eq:<H>-1-tibaduiza} and \eqref{eq:<H>-2-tibaduiza}] and for the mean number of excitations [Eqs. \eqref{eq:<N>-1-tibaduiza} and \eqref{eq:<N>-2-tibaduiza}] (which were not calculated in Ref. \cite{Tibaduiza-BJP-2020})
, and for the transition probabilities considering the initial state different from the fundamental [Eqs. \eqref{eq:Prob-m-n-r-Tibaduiza-1} and \eqref{eq:Prob-m-n-r-Tibaduiza-2}] (which generalizes the formula found in Ref. \cite{Tibaduiza-BJP-2020}).

We found that, as expected, the mean energy of the system is independent of time in each one of the intervals: $ t < 0$, $ 0<t < \tau $, and $ t>\tau $.
Moreover, we showed that the mean energy of the oscillator after the jumps is equal or greater than that before these jumps, even when $\omega_1<\omega_0$.
We also obtained, for $t>\tau\neq\tau_l$, a non-null value for the mean number of excitation when the oscillator starts in the fundamental state [Eqs. \eqref{eq:<H>-1-tibaduiza} and \eqref{eq:<H>-2-tibaduiza} with $ n=0 $], which means that, under the jumps in its frequency, a quantum oscillator, initially in the ground state, can become excited.
We showed that transitions between arbitrary $ m $ and $ n $ states only occur if $ |n-m| $ is an even number.
We highlighted that, for $t>\tau\neq \tau_l$ and a fixed value of $m$, as $n$ increases, $ {\cal P}(t)_{m\to n} $ decreases.
Finally, we showed that, for $t>\tau=\tau_l$, $ {\cal P}(t)_{m\to n}=\delta_{m,n}$, so that 
the oscillator returns to the same initial state (this generalizes, for any
initial state $m$, the result found in Ref. \cite{Tibaduiza-BJP-2020} 
for $m=0$).

\begin{acknowledgments}
	
The authors thank Alexandre Costa and Edson Nogueira for valuable discussions, as well as Adolfo del Campo, Bogdan M. Mihalcea, Daniel Tibaduiza, and Viktor Dodonov for their valuable suggestions to this paper.
S.S.C. was partialy supported by Conselho Nacional de Desenvolvimento Científico e Tecnológico - Brazil (CNPq) by the program PIBIC/CNPq through the project No. 144456/2020-6, Fundação Amazônia de Amparo a Estudos e Pesquisas (Fapespa) by the program PIBIC/Fapespa, and Coordenação de Aperfeiçoamento de Pessoal de Nível Superior - Brazil (CAPES), Finance Code 001. L.Q. was also supported by the CAPES, Finance Code 001.
	
\end{acknowledgments}
%


%

\end{document}